# Large orbital magnetic moments of small, free cobalt cluster ions Co$_n^+$ with $n \leq 9$

V Zamudio-Bayer[1,2,3], K Hirsch[2], A Langenberg[2], A Ławicki[2], A Terasaki[4], B von Issendorff[3] and J T Lau[1,3]

[1] Abteilung für Hochempfindliche Röntgenspektroskopie, Helmholtz-Zentrum Berlin für Materialien und Energie, Albert-Einstein-Straße 15, 12489 Berlin, Germany
[2] Institut für Methoden und Instrumentierung der Forschung mit Synchrotronstrahlung, Helmholtz-Zentrum Berlin für Materialien und Energie, Albert-Einstein-Straße 15, 12489 Berlin, Germany
[3] Physikalisches Institut, Universität Freiburg, Hermann-Herder-Straße 3, 79104 Freiburg, Germany
[4] Department of Chemistry, Kyushu University, 744 Motooka, Nishi-ku, Fukuoka 812-0395, Japan

E-mail: `vicente.zamudio-bayer@helmholtz-berlin.de`, `tobias.lau@helmholtz-berlin.de`

**Abstract.** The size dependent electronic structure and separate spin and orbital magnetic moments of free Co$_n^+$ ($n = 4-9$) cluster ions have been investigated by x-ray absorption and x-ray magnetic circular dichroism spectroscopy in a cryogenic ion trap. A very large orbital magnetic moment of $1.4 \pm 0.1$ $\mu_\text{B}$ per atom was determined for Co$_5^+$, which is one order of magnitude larger than in the bulk metal. Large orbital magnetic moments per atom of $\approx 1$ $\mu_B$ were also determined for Co$_4^+$, Co$_6^+$, and Co$_8^+$. The orbital contribution to the total magnetic moment shows a non-monotonic cluster size dependence: The orbital contribution increases from a local minimum at $n = 2$ to a local maximum at $n = 5$ and then decreases with increasing cluster size. The $3d$ spin magnetic moment per atom is nearly constant and is solely defined by the number of $3d$ holes which shows that the $3d$ majority spin states are fully occupied, that is, $3d$ hole spin polarization is 100%.

**Introduction**

One of the main motivations in cluster science is the investigation of how material properties develop from the atom to the bulk[1, 2] and one such material property that plays a fundamental role in modern day technology is the magnetization. Especially cobalt has attracted much attention in this context because of the relatively large orbital magnetic moment of the bulk material due to the hcp crystal structure.[3] The cluster size dependence of the total magnetic moment of free cobalt clusters has been successfully investigated in Stern-Gerlach deflection experiments.[4, 5, 6, 7] It was observed that the magnetic moment per atom in clusters with less than $\approx 500$ atoms is enhanced over the bulk value, increases with decreasing inverse cluster radius, and depends on the geometrical shell structure of the clusters. The enhancement of the magnetic moment is an expected consequence of the reduced size and the larger surface-to-volume ratio in clusters, both which lead to narrower bands and augment the spin imbalance. Additionally, orbital magnetic contributions emerge in small cobalt clusters, because they are less efficiently quenched than in the bulk material.[4] While large clusters have been intensely studied, the size dependence of the magnetic moment of metal clusters in the few atom size regime is still largely uncharted as with decreasing cluster size the Stern-Gerlach deflection experiment becomes more challenging. At the smallest sizes, the cluster beam intensities decrease rapidly and thermalization is more difficult to achieve.

More recently,[8, 9] x-ray absorption and magnetic circular dichroism spectroscopy (XAS and XMCD) of size-selected trapped ionic clusters have contributed to the study of cluster magnetism in a complementary way as the use of a buffer gas loaded cryogenic ion trap[9, 10] overcomes the difficulties of thermalization. Additionally, the application of the XMCD sum rules[11, 12] allows for the separation of spin and orbital contributions to the magnetic moment. We have now applied this technique to clusters with $n = 4 - 9$ atoms in order to test conflicting results from different Stern-Gerlach experiments[4, 5, 6, 7] at $n \lesssim 35$ and to deliver benchmark values for theoretical modeling of the orbital magnetic moment.[13, 14, 15] The interest in such modeling is strongly motivated by the pursuit of a better understanding of the magnetic anisotropy energy, which is of great technological importance and is closely related to the orbital magnetic moment.[16] Very recently, electronic structure calculations[13] have been performed on $Co_n$ clusters with the objective of overcoming the systematic underestimation of $\mu_L$ inherent to relativistic DFT calculations that were reported up to now.[17, 14] The most recent approach has included the orbital dependence of the on-site Coulomb interactions.[13, 15] In those calculations, the orbital magnetic moments per atom, $\mu_L \leq 0.7\ \mu_B$, determined by XMCD, of small $Co_n^+$ clusters with $n \geq 8$ and of the cobalt diatomic molecular cation ($\mu_L = 1\ \mu_B$) were already used as a benchmark.[18, 10, 8]

While in the meantime $\mu_s$ and $\mu_l$ from XMCD measurements have also been reported for the cobalt trimer cation[19] with $\mu_L \leq 0.5\ \mu_B$ per atom, accurate experimental values for the cluster size range from $n = 4 - 9$, where 3D cluster structures appear, are still missing. In the case of iron, already the smallest clusters have largely quenched orbital magnetic moments.[9] In the following, we present XMCD data that closes this gap from $n = 4$ to $n = 9$ and thus allows, through combination with Stern-Gerlach data on the neutral species, for a complete view of how the magnetic moment, with spin and orbital contributions resolved, develops from the atom to the bulk.

**Experimental Methods**

The $2p$ x-ray absorption spectra of free cobalt cluster cations were recorded at the NanoclusterTrap end station located at the UE52-PGM beamline of the BESSY II synchrotron radiation storage ring operated by Helmholtz-Zentrum Berlin. The setup has been described in detail before[20, 9, 10] and only a short summary will be given here. A liquid nitrogen cooled magnetron-sputtering gas-aggregation cluster source was used to produce a beam of cobalt cluster cations with a broad and tunable size distribution. Subsequently, the cluster size of interest was selected out of the ion beam with a quadrupole mass filter. The mass-filtered ion beam was then fed into the cryogenic ion trap, filled up to the space charge limit of $\approx 5 \times 10^7$ ions per cm$^3$. The cluster cations entering the ion trap are thermalized by the helium buffer gas to a temperature of $T \approx 15$ K and were magnetized in the magnetic field of $\mu_0 H = 5$ T generated by a superconducting solenoid. The high helium buffer gas pressure in the ion trap, in the order of $10^{-4}$ mbar, guarantees that many collisions between trapped ions and cold buffer gas atoms take place. This leads to thermalization and magnetization times in the order of ms while the lifetime of trapped ions is in the order of tens of seconds. The thermalized, trapped ion cloud was irradiated for every data point in the x-ray absorption and XMCD spectra for a period of 8 s with the monochromatic ($\Delta E = 625$ meV), circularly polarized (90% polarization) x-ray beam generated by the $3^{rd}$ harmonic of the undulator. The absorption of an x-ray photon by a parent cluster cation creates a core-hole and its relaxation induces the fragmentation of the absorbing cluster into smaller ions. The produced ion yield was analyzed using time-of-flight mass spectrometry while scanning the photon energy over the cobalt $L_{2,3}$ edge in 250 meV steps. The proportionality of the partial ion yield of the dominating Co$^+$ fragment channel to the total ion yield was verified by comparing the spectra resulting from different fragmentation channels in the resonant excitation region.[20]

A set of at least three x-ray absorption spectra was recorded each for right ($\sigma^+$) and left ($\sigma^-$) circular polarization of x-rays, respectively, per cluster size. The observed XMCD effect was quantified into spin and orbital magnetization with the help of the XMCD sum rules[11, 12]. The XMCD sum rules are explicitly given by:

$$m_S = -2 h_d \mu_B \left( \frac{3 \int_{L_3} (\sigma^+ - \sigma^-) - 2 \int_{L_{2,3}} (\sigma^+ - \sigma^-)}{\int_{L_{2,3}} (\sigma^+ + \sigma^-)} + \frac{7}{2} T_z \right)$$

$$m_L = -\frac{4}{3} h_d \mu_B \frac{\int_{L_{2,3}} (\sigma^+ - \sigma^-)}{\int_{L_{2,3}} (\sigma^+ + \sigma^-)}$$

with the number of 3d holes $h_d$, the spin magnetic dipole term $T_z$, and the approximation that the linear x-ray absorption cross section perpendicular to magnetization direction $\sigma^0$ is equal to the average of left and right circularly polarized x-ray absorption, $\frac{1}{2}(\sigma^+ + \sigma^-)$. This approximation has been shown to be valid in several cases, e.g. bulk iron and cobalt [21] and free clusters [9, 10]. In our experiment the magnetic field is antiparallel to the $k$-vector of the x-ray photons. Since absolute photon energies were not calibrated precisely, all spectra shown here have been rigidly shifted by 2.65 eV to higher photon energy in order to match the calibrated photon energy of previously published data. [Hirsch09] Relative energy shifts between individual cluster sizes are accurate to within 100 meV.

**Results**

*X-ray Absorption and XMCD Spectroscopy*

The polarization averaged $\frac{1}{2}(\sigma^+ + \sigma^-)$ x-ray absorption and corresponding magnetic circular dichroism $(\sigma^+ - \sigma^-)$ spectra of $Co_n^+$ with $4 \leq n \leq 9$ are shown in figure 1. Relative energies are not affected by this shift. Absorption and dichroism spectra are displayed on the same vertical scale, albeit with an offset for clarity, making the magnitude of x-ray absorption and XMCD directly comparable. The $L_{3,2}$ lines in the absorption spectra are narrower[20, 22] than in spectra of bulk samples and multiplet effects are clearly visible, which cause the appearance of an additional feature at the high energy side of the $L_3$ line. This additional line becomes well separated for clusters with less than 7 atoms where also the multiplet features become sharper and a third feature emerges at a photon energy of about 783 eV, marked in figure 1 by a dashed line. This third feature may be due to transitions into a narrow $s$ band. The appearance of this feature coincides with the increase in symmetry of the proposed geometric structure from the capped tetragonal bipyramid ($C_{2v}$) of $Co_7^+$ to the tetragonal bipyramid ($D_{3d}$) of $Co_6^+$.[23] A differing structural motif may also be the reason for the observed broadening and change in the high energy shoulder of the $L_3$ line in the spectrum of $Co_9^+$ with respect to the spectra of $Co_n^+$ $n = 7$ and 8. Indeed, while the reported structures of $Co_7^+$ and $Co_8^+$ are geometrically related, singly- and doubly-capped tetragonal bipyramids,[23], a tricapped trigonal prism was proposed for $Co_9^+$.[24] The observed multiplet effects in the x-ray absorption spectra are an indication of a considerable degree of spatial localization of the 3d states in comparison to the bulk material.[20]

The XMCD spectra in figure 1 are all qualitatively similar, showing a large negative dichroism at the $L_3$ edge and smaller positive dichroism at the $L_2$ edge. Both $L_3$ and $L_2$ dichroism lines are clearly asymmetric with tails on the high energy side. As for the absorption spectra, the XMCD spectra of dimer and trimer cations are more structured because of multiplet effects but apart from that, no cluster size is distinguished by a significantly different XMCD pattern. Note that the feature at ≈ 783 eV observed in the x-ray absorption spectra does not show dichroism. However, the XMCD spectra of cobalt clusters with four and five atoms stand out because of their particularly weak dichroism at the $L_2$ edge. This correlates to the already mentioned observation of clear multiplet effects in the corresponding absorption spectra. According to the orbital sum rule this weak positive dichroism at the $L_2$ edge is a clear indication of an increased orbital contribution to the magnetic moment.[25] In the following the XMCD sum rules[11, 12] are used in order to quantify the orbital contribution to the total magnetic moment as a function of cluster size.

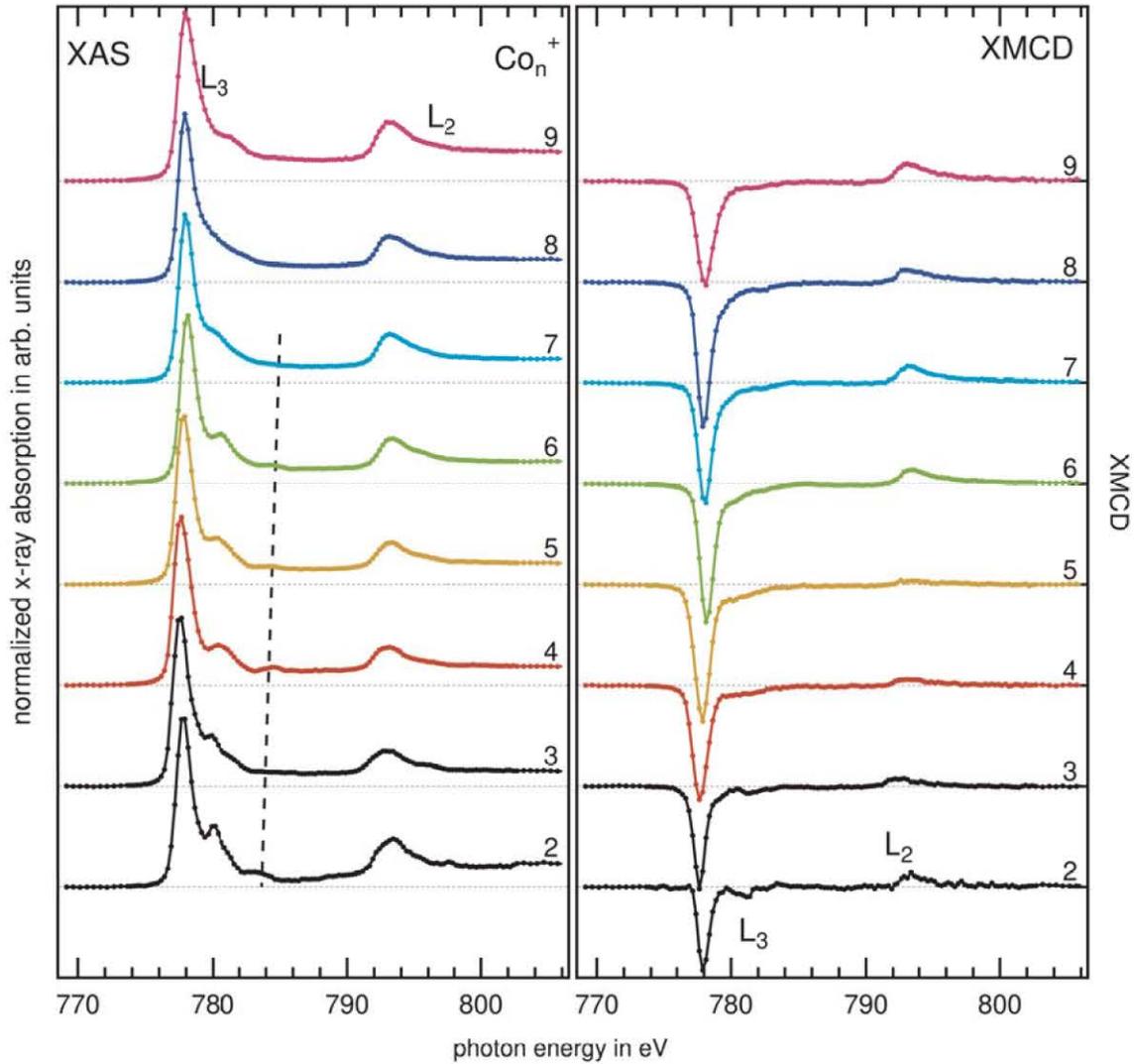

**Figure 1.** Cobalt 2p x-ray absorption average ($\frac{1}{2}(\sigma^+ + \sigma^-)$) of left and right circularly polarized light (left) and XMCD ($\sigma^+ - \sigma^-$) (right) spectra of $Co_n^-$ clusters ($n = 4 - 9$). For comparison, spectra of dimer[18] and trimer[19] cations are also shown. Spectra are shown on the same scale but vertically offset for clarity.

*Orbital-to-spin magnetic moment ratio*

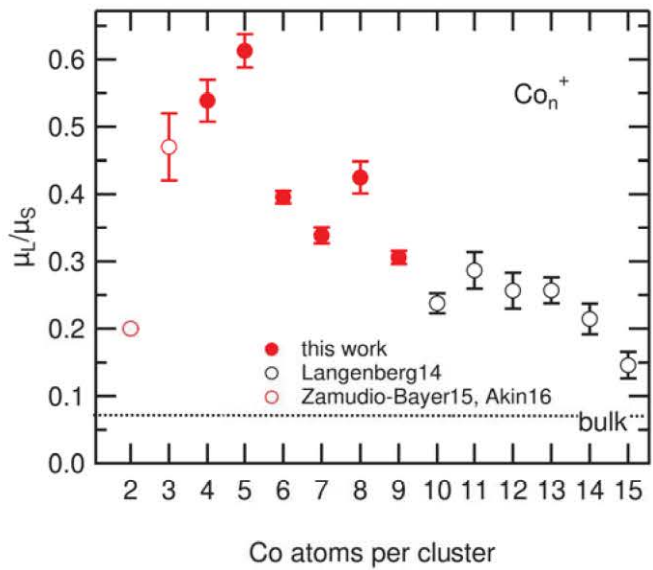

**Figure 2.** Experimentally determined orbital-to-spin magnetic moment ratio as a function of cluster size. For comparison the values for $Co_n^+$ clusters with $n = 2, 3, 10-15$ [18, 19, 10] and the value of 0.07 for bulk cobalt[3] are also shown. Note that the ratio for the cobalt cation in its $[Ar]3d^8\ ^3F$ ground state is 1.5 and for the cobalt atom with $[Ar]3d^74s^2\ ^4F$ ground state is 1, both of which are outside the scale that is shown here.

The ratio of orbital-to-effective-spin magnetic moment $\langle L_z \rangle / (\langle 2S_z \rangle + 7 \langle T_z \rangle)$ can be determined with high accuracy, because any dependence on the number of holes $h_d$, ion temperature $T_{ion}$, or isotropic spectrum $\sigma^0 + \sigma^+ + \sigma^-$ is canceled from the XMCD sum rules. The remaining spin magnetic dipole term $T_z$ is estimated to be negligible as it has been shown to be $\leq 10\%$ of the spin even in iron, cobalt, and nickel dimer cations[18, 26] where it would be expected to be of considerable magnitude. In figure 2 the orbital-to-spin magnetic moment ratio $\mu_L/\mu_S = L_z/2S_z$ obtained for $Co_n^+$ is shown as a function of the number of atoms per cluster $n$. It can be seen that the ratio $\mu_L/\mu_S$ is a non-monotonic function of the cluster size $n$, strongly increasing from $n = 2$ to $n = 5$ by almost a factor of three. With further increasing cluster size the $\mu_L/\mu_S$ ratio decreases with an additional local maximum at $n = 8$. Note that the small ratio for the dimer is not a consequence of symmetry-induced orbital magnetic moment quenching, as $L_z$ along the intramolecular axis ($\Lambda$) is still a good quantum number in the diatomic molecular ion. Instead, the small orbital magnetic moment in $Co_2^+$ is due to its electronic configuration with three electrons in a doubly degenerate orbital of $\pi$ symmetry with $m_l = \pm 1$.[18] The considerable difference between the orbital-to-spin magnetic moment ratio of $Co_{4,5}^+$ and $Co_6^+$ indicates an abrupt change in the symmetry of the $3d$ derived states between $n = 5$ and $n = 6$. Overall the ratio in the $n = 2 - 15$ cluster size range is considerably higher, by a factor of two to nine, than in bulk cobalt[3], as indicated in figure 2.

*Size dependent spin and orbital magnetic moments*

From the application of the XMCD sum rules to the experimental data shown in figure 1 we obtain the spin and orbital magnetization per cluster normalized to the number of unoccupied $3d$ states ($3d$ holes), at the experimental conditions of magnetic field strength $\mu_0 H$ and temperature $T_{\text{ion}}$, for each cluster size. The magnetization of the free clusters as a function of the magnetic field strength and cluster temperature is described in first-order approximation by the Brillouin function. It is therefore possible to obtain the saturation magnetization, that is, the magnetic moment $\mu_J$, using the Brillouin function. The Brillouin function has the parameters $h_d$, $T_{\text{ion}}$, $\mu_0 H$, and the $\mu_L/\mu_S$ ratio. As an initial assumption for the *a priori* not exactly known ion temperature and number of $3d$ holes, we resort to our previous experiments[18, 10] on $Co_n^+$ with $n = 2, 3, 10 - 15$ at comparable experimental parameters as the ones used in the present work. In our previous experiments[18, 10] we were able to show that the number of $3d$ holes is almost constant in this size range and equals the value of bulk cobalt.[21] The starting assumption is, therefore, a number of $3d$ holes $h_d = 2.5$ and an ion temperature $T_{ion} = 12 \pm 3$ K. Also from the same previous studies, we expect complete hole spin polarization. In case of full hole spin polarization (spin magnetic moment of 1 $\mu_B$ per hole), the hole spin magnetization obtained from the application of the spin sum rule to the measured XMCD spectra should be equal to the total (spin plus orbital) magnetization obtained from the Brillouin function because all holes are situated in the minority spin band. The comparison of hole spin magnetization from XMCD and expected Brillouin magnetization is shown in figure 3. Overall the measured hole spin magnetization for $Co_{6-9}^+$ matches the expectations. The less good agreement for $Co_4^+$ and $Co_5^+$ could be accounted for by slightly higher ion temperatures of 18 K and 23 K, respectively. Increased temperatures of the smallest clusters is not surprising as the smaller clusters can be expected to be more strongly affected by radio-frequency heating induced by the ion trap.[9] Other possible explanations would be a reduced number of $3d$ holes or a broadening of majority and minority spin bands leading to an overlap of both bands with the Fermi energy and thus to incomplete hole spin polarization. The spectroscopic indication of the former would be a reduced total absorption intensity, and the latter would lead to a markedly reduced $L_3$ to $L_2$ branching ratio. As we do not observe any clear spectroscopic indication of neither one, the higher cluster temperature is the most probable explanation for the deviation from the expected magnetization.

While the starting assumption of 2.5 holes in the $3d$-derived states is reasonable, a more detailed estimate of the number of $3d$ holes results from the expected electronic configu-ration of $\text{Co}_n^+$ clusters, $3d^{7n+1+2m}4s^{2(n-1-m)}$, with $m \leq n-1$, which considers a possible $4s$-$3d$ transfer of an even number $2m$ of electrons. We expect the number of occupied states in the $4s$ band to be even because in the few atom regime the density of states is still discrete and the $4s$ exchange interaction is too small to cause any considerable $4s$ spin imbalance. Thus, any additional atom for a given cluster size contributes either 1 $\mu_B$ or 3 $\mu_B$ to the spin magnetic moment. For $\text{Co}_{4-9}^+$ we can therefore conclude that the spin magnetic moment is determined by the number of $3d$ holes and that the hole spin polarization is 100 %. Applying these constraints we can assign the number of holes $h_d$ and the corresponding spin multiplicities and orbital magnetic moments for each cluster size as listed in table 1.
The corresponding magnetic moment per atom as a function of cluster size with spin and orbital contribution resolved is shown in figure 4. As the number of 3$d$ holes varies only slightly around 2.5 (see table 1), the spin magnetic moment per cobalt atom is close to constant with cluster size, in other words, the magnetic moment per cluster increases almost linearly with increasing cluster size. The full spin polarization of 1 $\mu_B$ per hole, i.e. , a spin magnetic moment per atom of $\approx 2.5$ $\mu_B$, follows the initial assumption that all the clusters studied here have all their atomic spins coupled in parallel fashion and are thus single domain, as was the case for $Co_{10-15}^+$.[10] The orbital magnetic moment per atom, while showing a general trend to decrease with cluster size, does not behave strictly monotonic as a function of cluster size. When comparing with the high $\mu_L/\mu_S$ ratio for $Co_{4,5}^+$ seen in figure 2 it is clear that this exceptionally high ratio is caused by a combination of both an increased orbital magnetic moment per atom and a lower spin magnetic moment per atom due to a slightly reduced number of 3$d$ holes.

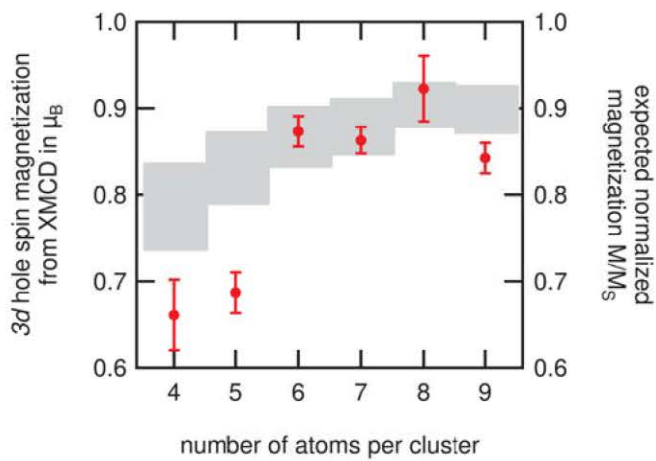

**Figure 3.** Measured $3d$ hole spin magnetization in $\mu_B$ (left) and expected normalized magnetization $M/M_S$ (right) for complete $3d$ hole spin polarization of $\text{Co}_n^+$ as a function of cluster size $n$. The boundaries of the expected magnetization (gray) result from the upper (15 K) and lower (9 K) boundaries on the ion temperature. The ion temperature is underestimated for $n = 4, 5$. (see text for a detailed discussion)

**Table 1.** Spin multiplicities $2S+1$ and electronic orbital angular momenta $L_z$ in $\hbar$ per cluster and per atom for $\text{Co}_{4-9}^+$. The number of $3d$ holes per atom $h_d$ is listed in the third column and is equal to the average spin magnetic moment per atom due to complete hole spin polarization.

|  | $2S+1$ | $L_z$ in $\hbar$ | $h_d$ per atom ($\equiv \mu_s$ per atom in $\mu_B$) | $\mu_L$ per atom in $\mu_B$ |
| --- | --- | --- | --- | --- |
| $\text{Co}_4^+$ | 10 | $4.8 \pm 0.2$ | 2.25 | $1.21 \pm 0.07$ |
| $\text{Co}_5^+$ | 13 | $7.4 \pm 0.3$ | 2.4 | $1.47 \pm 0.06$ |
| $\text{Co}_6^+$ | 16 | $5.9 \pm 0.3$ | 2.5 | $0.98 \pm 0.03$ |
| $\text{Co}_7^+$ | 19 | $6.1 \pm 0.4$ | 2.58 | $0.87 \pm 0.03$ |
| $\text{Co}_8^+$ | 20 | $8.1 \pm 0.5$ | 2.38 | $1.00 \pm 0.06$ |
| $\text{Co}_9^+$ | 23 | $6.7 \pm 0.5$ | 2.44 | $0.74 \pm 0.03$ |

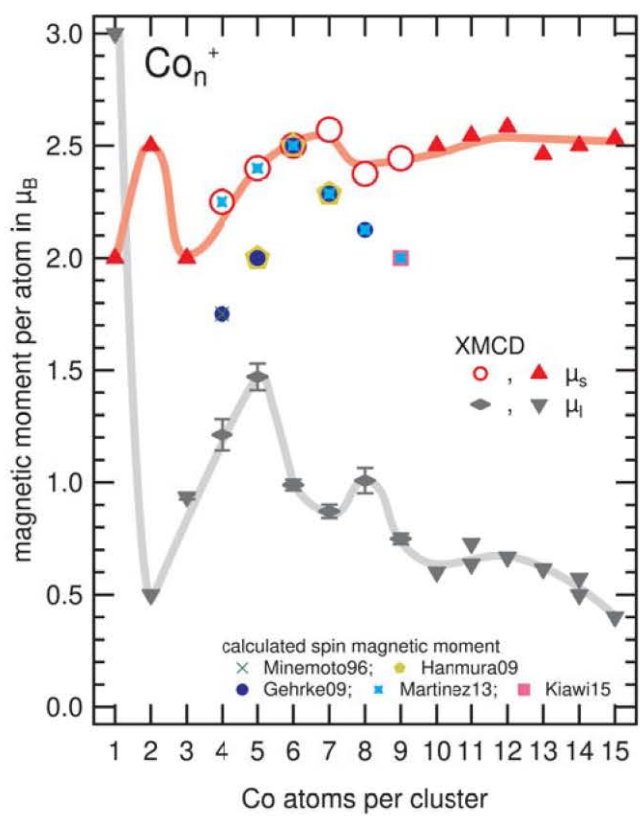

**Figure 4.** Spin (empty circles) and orbital (rhombi) magnetic moments per atom of $Co_{4-9}^+$ clusters determined with the help of XMCD sum rules (this work) and of $Co_{2,3,10-15}^+$ (triangles).[19, 10, 18]. Theoretical predictions by Minemoto96 [27] for $Co_4^+$ (cross), Hanmura09 [28] for $Co_{5-7}^+$ (pentagons), Gehrke09 [23] for $Co_{4-8}^+$ (full circles), Kiawi15 [24] for $Co_{6,9}^+$ (squares), and Martinez13 [29] for $Co_{4-9}^+$ (stars) are also shown. The lines are guides to the eye.

Theoretical predictions[23, 27, 28, 29, 24] for the spin magnetic moments for $Co_{4-9}^+$, available in the literature, are also shown in figure 4. Apart from reference [27] where the discrete variational $X\alpha$ (DV-$X\alpha$) method was used, all theoretical studies have used density functional theory (DFT) with the generalized gradient approximation (GGA) to the exchange-correlation functional. When comparing theoretical predictions and experimental results of the present study, it is worth noting that only for $Co_6^+$ all predictions and our experimental results agree on a multiplicity of $2S+1 = 16$, that is, $\mu_S = 15\ \mu_B$. Overall, the predictions systematically underestimate the spin magnetic moment, a trend which has already been observed in the $n = 10-15$ size range for $Fe_n^+$, $Co_n^+$, and $Ni_n^+$ clusters.[10] Nevertheless, a qualitatively similar evolution of the spin magnetic moment as a function of cluster size can be observed in the experimental values and theoretical predictions, with a local maximum at $n = 6, 7$. When comparing with predictions for the neutral species, not shown in the figure, a correlation between cluster size and effect of electron removal on the spin multiplicity can be observed. A common assumption is that the spin multiplicity of ionic and neutral species differs by $\pm 1$ because ionization or electron attachment are considered one-electron processes. For the smaller $Co_{4,5}$ clusters, the comparison to the predictions for neutrals, not shown in figure 4, shows that almost all studies predict ground state spin multiplicities for the neutral species which are higher than the ones determined in this work, and also higher than predicted,[23, 27, 28, 29] for the cationic species.[13, 30, 31, 32, 33, 34, 35, 36, 37, 38, 29, 39, 40, 41, 42] Contrastingly, for the larger neutral $Co_{6-9}$ clusters, the majority[32, 33, 34, 35, 36, 37, 38, 29, 40, 41] of the studies predict lower spin multiplicities for the neutral species than the ones determined here for the cationic ones, and even lower than predicted for cations,[23, 28, 29, 24] while only two studies[13, 30] predict higher spin multiplicities.

For the orbital magnetic moment of $Co_n^+$ with $4 \leq n \leq 9$ there are no predictions available in the literature. For the neutral species, a recent theoretical study[13] predicted $\mu_L = 0.67-0.74\ \mu_B$ per atom for $Co_{4-9}$ clusters, in reasonable agreement with experimental results for cationic clusters $Co_n^+$ with $n = 2, 8, 9$ available at the time.[8, 18] It should be noted that orbital magnetic moments $\geq 0.5\ \mu_B$ per atom are only predicted by theory when explicitly taking into account the orbital dependence of the on-site Coulomb interactions.[15, 13] In contrast to theoretical predictions for neutral clusters[13], the results presented here indicate that the orbital magnetic moment per atom has a strong size-dependence, strongly increasing with increasing cluster size for $Co_n^+$ $2 \leq n \leq 5$ and decreasing for clusters with $n \geq 6$.

## Interpretation & Discussion

*High orbital-to-spin magnetic moment ratio, geometric structure, and reactivity*

The exceptionally high orbital-to-spin magnetic moment ratio $\mu_L/\mu_S$ observed for $Co_{4,5}^+$ coincides with their reported increased reactivity towards $H_2$, $N_2$, $CH_4$, and $C_2H_4$. [43] The increased reactivity at these cluster sizes is believed to be caused mainly by geometric constraints but not by the specific electronic structure because it has been observed for various transition metal clusters independent of their specific filling of the 3d shell.[43] As an explanation it was proposed that the non-metal reagents are activated by donated electrons from the metal cluster. This implies that the geometric structures of cationic TM clusters with $n = 4, 5$ favor the donation of electrons. Electrons easily available for donation should not be crucial to the metal-metal bonding in the cluster but are instead non-bonding and localized at their atomic sites. The symmetry at the site of these localized electrons would be less perturbed by the surrounding electronic density and a considerable amount of orbital angular momentum could survive due to the, at least partially, preserved orbital degeneracy. The degree of localization and thus the amount of surviving orbital magnetic moment then strongly depends on the specific geometric structure. This reasoning is in agreement with the observed non-monotonic cluster size-dependence of the orbital magnetic moment $\mu_L$ per atom, as shown in figure 4.

*Orbital magnetic moment and magnetic anisotropy energy*

The observed large orbital magnetic moment of 1.4 $\mu_B$ per atom for $Co_5^+$ is much larger than the highest predicted values for small cobalt clusters of up to 0.86 $\mu_B$ per atom.[13] It is worth noting that average values of $L_z \geq 1$ $\mu_B$ per atom from 3d orbitals require incomplete occupation of degenerate orbitals with both $m_L = \pm 1$ and $m_L = \pm 2$. On metallic surfaces, only for the cobalt adatom on Pt(111) a comparable large value of $1.1 \pm 0.1$ $\mu_B$ per cobalt atom has been measured so far.[44] For this system, a magnetic anisotropy energy (MAE) of 9 meV per cobalt atom was determined and a strong correlation of the magnitude of the MAE to the magnitude of orbital angular momentum $L$ was found. It was also observed that, while $L_z$ reacts very sensitively to changes in the local coordination, the spin angular momentum $S$ does not. The robustness of the spin $S$ is due to a nearly filled majority spin band. The strong dependence of $L_z$ on the local environment has also been predicted by theory for small, free transition metal clusters in calculations, where the orbital dependence of the intra-atomic Coulomb interactions was taken into account.[15] Both observations of filled majority spin states and a strong dependence of $L_z$ on the geometric structure, are in very good agreement with the cluster size-dependent behavior described in the present work. Because a filled majority spin band has been determined with XMCD for cobalt clusters with 10 to 15 atoms before,[10] any opening of the majority spin band due to, for example, an increased overlap with the spin minority band and crossing of the Fermi energy should occur at larger sizes.

For cobalt clusters on surfaces[44], $L_z$ decreased rapidly with increasing cluster size and the remaining MAE was then predominantly due to the large spin-orbit coupling of the Pt surface. In our case, the large $L_z$ is intrinsic to the $Co_5^+$ cluster as there is no interaction with any surface but the predicted bipyramid geometry[23, 28, 29] places all atoms at vertices, with the symmetry at the local sites far from the four-fold rotational point symmetry that would lead to a quenching of the orbital angular momentum. In $Co_5^+$ low coordination leads to $d$-band narrowing and thus to an increased spin-orbit coupling (SOC) energy.[45, 44, 46] The enlarged SOC energy could lead to a considerable MAE, much higher than the $\approx 60$ $\mu$eV per atom of bulk cobalt[47] and possibly even comparable to the 9 meV per cobalt atom on platinum surfaces.[44] A very rough estimate of the MAE of $Co_5^+$ can be estimated from the Bruno model[45] with the $3d$ SOC energy of cobalt, $\xi \approx 64$ meV and the large orbital magnetic moment of 1.4 $\mu_B$. [48] We estimate the orbital magnetization in the hard-plane $L_{xy}$ of $Co_5^+$ to be comparable in magnitude to the average orbital magnetization $L_z = 0.9$ $\mu_B$ of $Co_n^+$ with $n = 4, 6$–$9$. One then indeed obtains[45] $64 \, \text{meV}/4 \times (1.4(\text{easy}-\text{axis}) - 0.9(\text{hard}-\text{plane})) \approx 8$ meV per Co atom for $Co_5^+$.

In the bulk, the MAE is typically associated with a blocking temperature below which the magnetic moment locks onto the fixed easy-axis of magnetization. The situation is opposite for free clusters, where the magnetic moment still precesses around the applied magnetic field but the lattice follows this precession, leading to hindered rotation and thus to spatial alignment. The corresponding blocking temperature in our experiment, according to the Arrhenius relation and due to the lifetime of clusters in the ion trap in the order of $\approx 10$ s, would then equal $5 \times 8 \, \text{meV} \times (k_B \ln{(10 \, \text{s}/10 \, \text{ps})})^{-1} \approx 17$ K. The attempt period of 10 ps is estimated from the Larmor frequency at 5 T.

The expected effective degree of spatial alignment of the clusters achieved at the current temperature is low due to the high density of rotational states and thus no considerable spectroscopic evidence of a net spatial alignment in the trap can be expected. Nevertheless, it only introduces a small error in the magnetic moment since, as we have shown previously, even for the extreme case of cobalt diatomic molecular cations with higher MAE, lower density of rotational states, and cooled to lower temperature, the deviation from the Brillouin function is small.[18, 26] An improved description would require an effective Zeeman-Hamiltonian.[49, 50, 18] Remarkably, some indications of emerging blocked-moment behavior has been observed in Stern-Gerlach experiments on small Co clusters at temperatures around 50 K.[5, 6] In summary, we expect the trigonal bipyramid $Co_5^+$ cluster to have a considerable MAE due to its large orbital magnetic moment and its elongated structure, but significantly lower ion temperatures are required for spectroscopic evidence of blocking.

*Total magnetic moments from atom to bulk*

Combining the values obtained here for the total magnetic moment with values from Stern-Gerlach experiments and from our previous XMCD experiments, we can now plot the magnitude of the total magnetic moment per atom $\mu_J$ as a function of inverse cluster radius over the whole size range from the dimer to clusters containing hundreds of atoms. This is shown in figure 5. It should be noted though, that although there is no obvious reason why the charge on the cations should make much of a difference in clusters with

> 100 valence electrons, it cannot be strictly ruled out as exemplified by the peculiar case of $Fe_{13}^+$.[9, 10] In the smallest cobalt clusters with $n \leq 10$ the magnitude of the total magnetic moment develops non-monotonically as cluster size increases, depending on the particular geometry of the cluster. In this size range, the spin magnetic moment oscillates between $2 - 2.58$ $\mu_B$ per atom and the orbital contribution to the total magnetic moment is significant. For $10 \lesssim n \lesssim 20$, the spin magnetic moment of $\approx 2.5$ $\mu_B$ per atom is close to constant while the orbital magnetic moment decreases with increasing cluster size. For larger sizes $35 \lesssim n \lesssim 500$ the orbital magnetic moment is strongly suppressed and the total magnetic moment decreases almost linearly with decreasing inverse cluster radius due to the increasing broadening and overlap of both, majority and minority states with the Fermi energy, which leads to the opening of holes in the majority spin band. The total magnetic moment levels off at $n \approx 500$, where it reaches the value of bulk hcp cobalt. The decreasing surface-to-volume ratio ($S/V \propto R^2/R^3 = 1/R$) leads to an increase of the average atomic coordination and to an accompanying $3d$ band broadening resulting in a reduction of spin imbalance. The onset of the reduction of spin imbalance by increasing width of minority and majority bands can be estimated to occur at the latest at cluster size $n \approx 37$ where the total magnetic moment per atom drops below the average number of $3d$ holes of $\approx 2.5$ per atom. Around $n \approx 37$ is also the smallest cluster size at which the results of the different Stern-Gerlach experiments all agree within experimental uncertainties. For clusters with $n \leq 25$ the Stern-Gerlach experiment is more challenging than for larger sizes due to thermalization issues, a fact that is mirrored by the disagreement between the different reported results for cluster sizes down to $n = 13$.

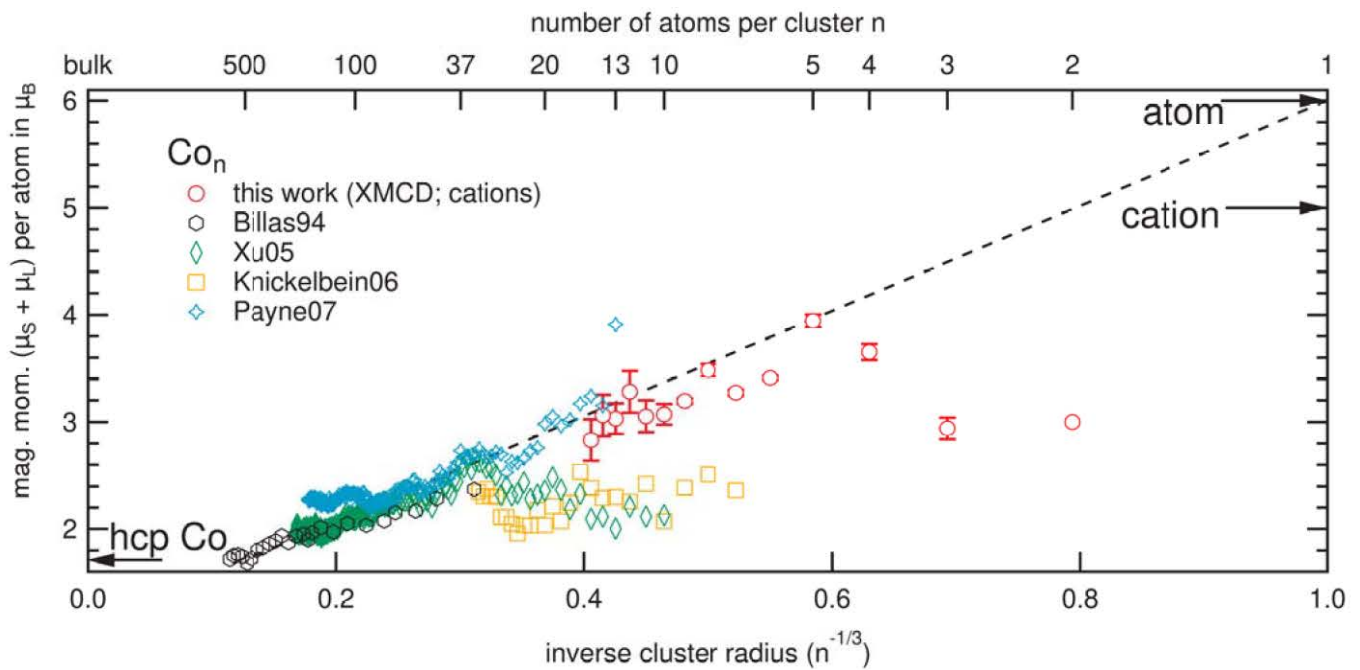

**Figure 5.** Total magnetic moment per atom of $Co_n$ clusters as a function of the inverse cluster radius. The values for $Co_n^+$ with $n = 2, 3, 10-15$ are taken from our previous XMCD results.[10, 18, 19] Results from Stern-Gerlach experiments on neutral species available in the literature, Billas94 [4], Xu05 [5], Knickelbein06 [6], Payne07 [7] are shown for comparison. Also shown are the values for hcp cobalt[3] and for the atom and ion in their corresponding $^4F$ and $^3F$ ground states, respectively. For clusters with $5 \leq n \leq 500$, the total magnetic moment roughly follows a $1/R$ dependence, as indicated by the broken line. See text for a discussion of the orbital and spin contributions.

Examining now the smallest clusters and beginning with the dimer, the comparison with figure 2 clearly shows that bond formation strongly reduces the orbital contribution to the total magnetic moment. The reduction is especially dramatic in the case of cobalt as both the atom and cation have an orbital magnetic moment of 3 $\mu_B$ which is the largest orbital magnetic moment of a $3d$ element in its atomic or ionic ground state, according to Hund's rules. From figure 2 we know that the orbital magnetic moment strongly decreases from the atom to the dimer, then increases for $n = 2-5$ and then decreases again with cluster size for $n \geq 6$. It is already reduced by a factor of $\approx 5$ from its atomic value at $n = 9$, and by a factor of $\approx 10$ at $n = 15$. As we have shown here, the spin magnetic moment per atom is close to constant while the orbital magnetic moment strongly depends on the exact cluster size and geometry in smaller clusters with $n \leq 15$, corresponding to the size regime of emergent phenomena, and thus a simple scaling law[51] cannot be expected to hold. If at all, the use of scaling laws for spin or orbital magnetic moments may be valid for $40 \lesssim n \lesssim 500$.

**Conclusions**

In conclusion, we find that for $Co_n^+$ clusters the orbital contribution to the total magnetic moment is highest in the $3 < n < 10$ cluster size range and even dominates the variation of the total magnetic moment for $2 > n > 5$. We also find that the spin magnetic moment per atom is close to constant at least up to $n = 15$ as the hole spin polarization in small cationic cobalt clusters is unity. The magnitude of the spin magnetic moment per cluster in this size range is thus determined by the number of $3d$ holes per atom, which does not vary in a considerable way as a function of cluster size. $Co_{4-9}^+$ clusters have a large orbital magnetic moment per atom of up to 1.4 $\mu_B$ for $Co_5^+$, which is one order of magnitude larger than in bulk hcp cobalt. This exceptionally large orbital magnetic moment of $Co_5^+$ is a sign of the highly localized nature of the $3d$ orbitals in this particular cluster. A previously reported maximum in the reactivity at this cluster size probably is the effect of this strong $3d$ orbital localization. Furthermore, the $Co_5^+$ cluster possibly possesses a large magnetoanisotropy energy, which we estimate to be of the order of 10 meV per atom, much larger than most $3d$ transition metal-based single-molecule magnets and comparable to values obtained for lanthanide-based ones.[52] Lastly, we find that in clusters with less than ten atoms, the strong dependence of the orbital magnetic moment on the specific cluster geometry, together with the dependence of the spin magnetic moment on the number of $3d$ holes, dominate over any simple dependence on the inverse cluster radius. This size regime is thus non-scalable where emergent phenomena are observed but no simple scaling law can be expected to be valid for spin or orbital magnetic moments.

## Acknowledgments

We thank HZB for the allocation of synchrotron radiation beamtime at the beamline UE52-PGM. The superconducting magnet was provided by Toyota Technological Institute. AT acknowledges financial support by Genesis Research Institute, Inc. This project was partially funded by the German Federal Ministry for Education and Research (BMBF) under Grant No. BMBF-05K13VF2 and No. BMBF-05K16VF1. BvI acknowledges travel support by HZB.